\documentstyle[proceedings,epsfig]{crckapb}
\begin{opening}
\title{Dynamics of the asymmetries \protect\\
       at galactic centers}
\author{F. Masset}
\author{M. Tagger}
\institute{Service d'Astrophysique (CNRS URA 2052), CEA-Saclay, 91191
Gif/Yvette Cedex, France}
\end{opening}
\begin{document}
\section{Introduction}

Bars in galaxies are frequently found to be off-centered, so that the
center of
the bar does not coincide with the center of rotation, with a difference which
typically amounts to several tens of parsecs.  This is clearly seen in the
distribution of the gas and the rotation curves at the center of the Milky
Way,
and in nearby barred galaxies (Blitz, these proceedings, and del Burgo et al.,
these proceedings).

We propose here a new dynamical mechanism to account for the bar
off-centering.
It is based on the presence of an $m=1$ density wave, whose first
manifestation will be the off-centering of the central region of the galaxy
(see
also Miller and Smith, 1992, Combes, these proceedings, and Junqueira and
Combes, 1996).  The mechanism we propose is a non-linear
excitation of the $m=1$ perturbation by the strong $m=2$ due to the linearly
unstable bar and $m=3$ mode.

\section{Non-linear coupling}
If two spiral waves coexist in a disk, we expect them to excite beat waves at
the sum and difference frequencies and wavenumbers: thus in particular an
$m=2$
and an $m=3$ spirals, with frequencies $\omega_{2}$ and $\omega_{3}$ create a
beat wave with $m=1$ and $\omega_{1}=\omega_{3}-\omega_{2}$.  If at this
frequency and wavenumber the beat wave obeys the dispersion relation, i.e.
can
propagate in the disk, it can very efficiently exchange energy and angular
momentum with the parent waves.  We have already found (see Masset and Tagger,
1997a, and references therein), from analytical and numerical work, a
number of
examples where this non-linear coupling can play an important role in the
evolution of spirals and warps.  Thus we expect that, in disks where the $m=2$
and $m=3$ spirals are linearly unstable by the classical Swing mechanism, they
can excite an $m=1$ wave which could explain the off-centering of the central
region of the disk.  In order to check this, we have performed numerical
simulations with a polar Particle-Mesh code, whose small grid size in the
central region allows precise physics and diagnostics.

\section{Results}

\medskip

\psfig{file=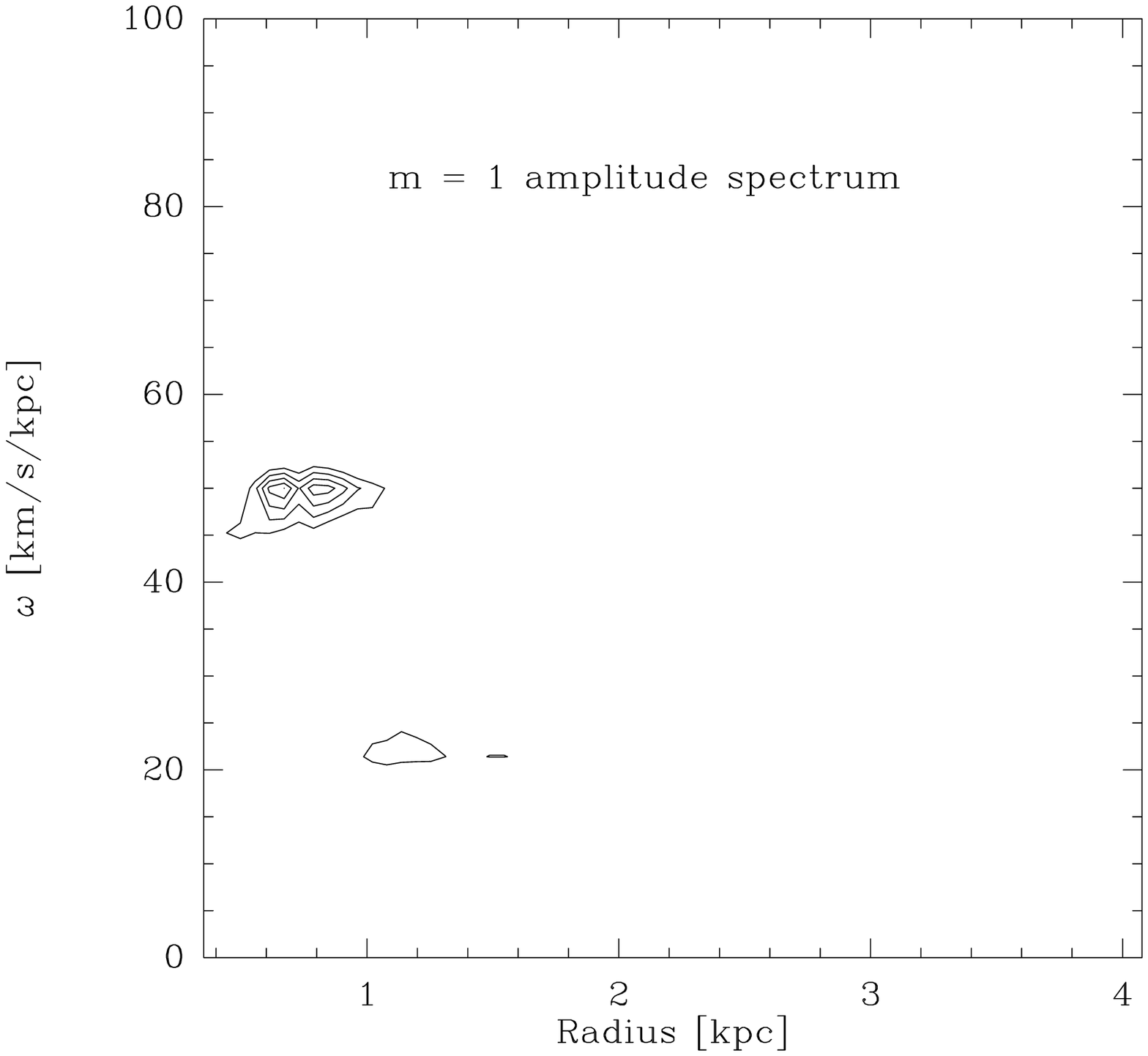,width=4.2cm}\psfig{file=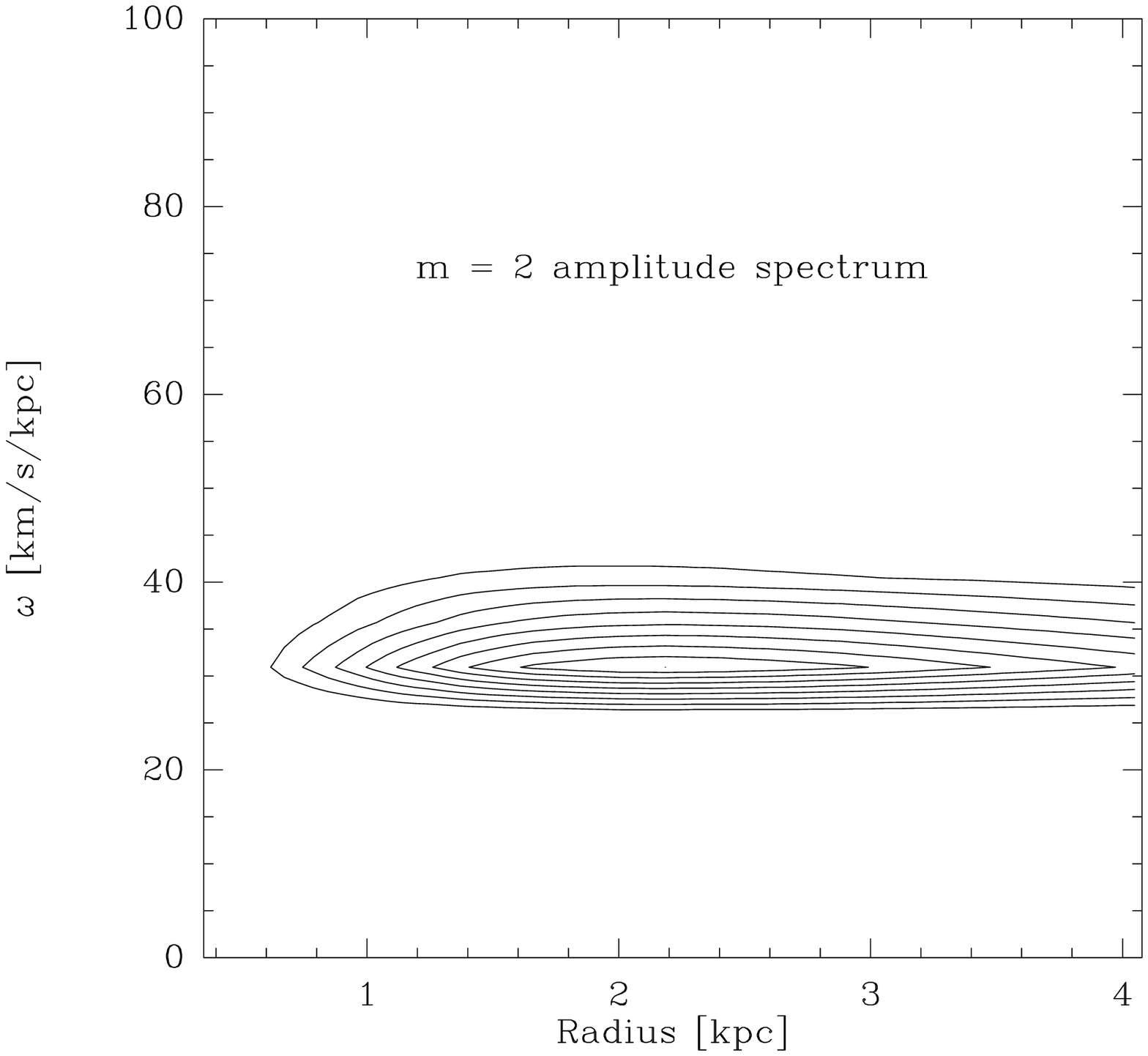,width=4.2cm}\psfig{file=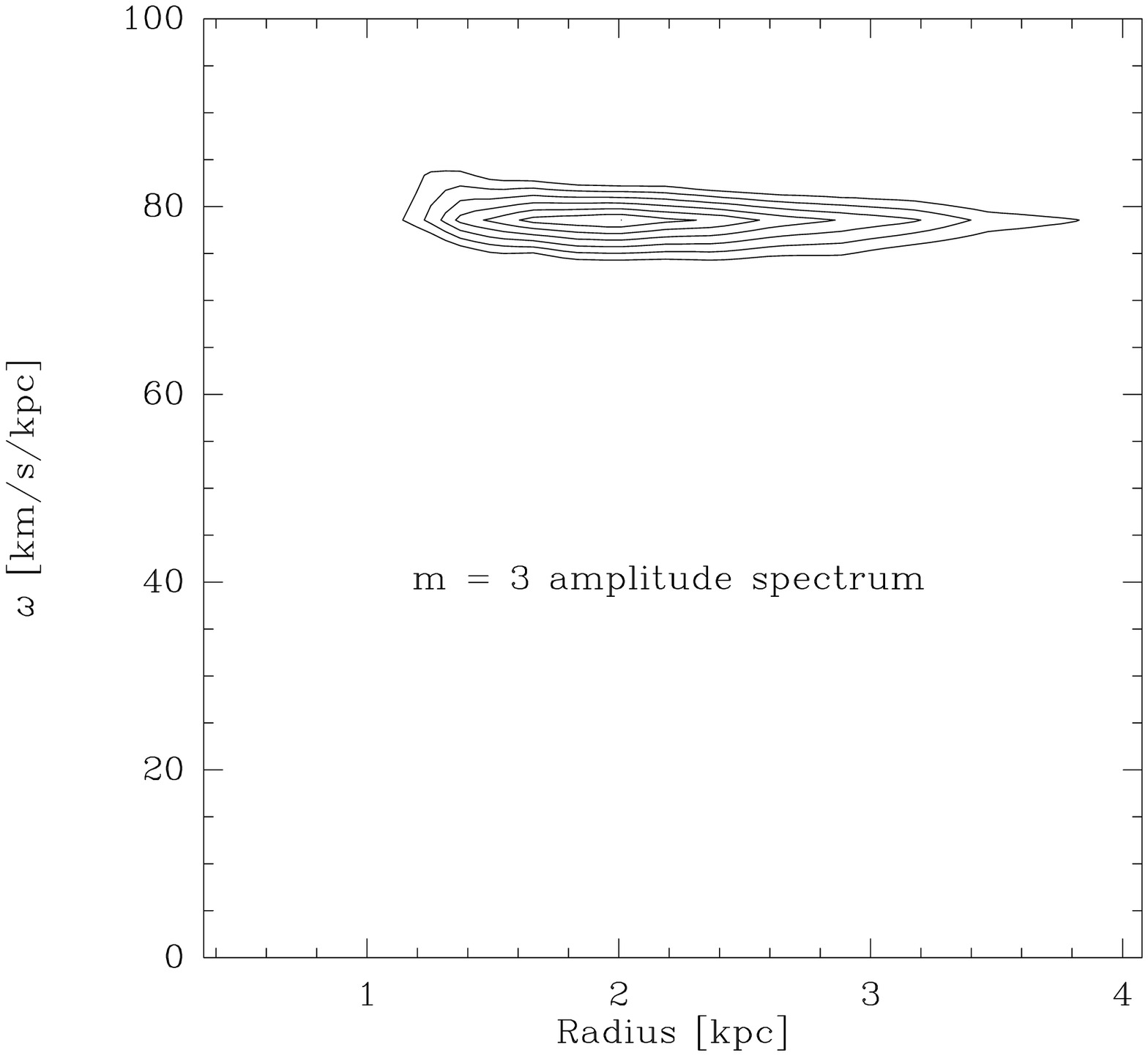,width=4.2cm}

\medskip

We show above the amplitude spectra (Masset and Tagger, 1997b) of the $m=1$,
$m=2$ and $m=3$ relative perturbed densities for a typical simulation.  We see
the signature of the bar on the $m=2$ plot, at the frequency
$\omega_2=30$~km/s/kpc; on the $m=3$ we see a well defined mode at
$\omega_3=80$~km/s/kpc, and on the $m=1$~plot the expected beat wave at
$\omega_1=80-30=50$~km/s/kpc.  This wave corresponds to
a~60~pc deviation of the center of gravity of the inner regions.  In order to
specify the energy transfer between waves we have redone the simulation by
setting
to zero at each timestep the $m=1$ potential component, and hence inhibiting
possible $m=1$ linear modes.  We still see the $m=3$ mode, which proves
that it
is not excited by the non-linear coupling.  When we redo the simulation by
setting to zero the $m=3$ potential component, we do not see
the $m=1$: this shows that it needs the other modes to be excited, and
hence that it is non-linearly excited by the linearly unstable bar and $m=3$
modes.

As a conclusion,
this mechanism
naturally leads to an $m=1$ perturbation (i.e. an off-centering) whose
amplitude is sufficient to explain the asymmetries typically observed.
Work is in progress to include self-gravitating gas
and a live bulge in the simulations, for a more detailed description.

\bigskip

\parindent=0mm

{\bf {\large {\bf References}}}

\bigskip

\small
Junqueira, S. and Combes, F., 1996 : A\&A, 312, 703

Masset, F. and Tagger, M., 1997a : A\&A, 318, 747

Masset, F. and Tagger, M., 1997b : A\&A, 322, 442.

Miller, R.H. and Smith, B.F., 1992 : ApJ, 393,508

\end{document}